# Dual-Domain Reconstruction Networks with V-Net and K-Net for fast MRI

Xiaohan Liu[1], Yanwei Pang[1], Ruiqi Jin[1], Yu Liu[1], Zhenchang Wang[2]

[1]Tianjin Key Lab. of Brain Inspired Intelligence Technology, School of Electrical and Information Engineering, Tianjin University, Tianjin, China

[2]Beijing Friendship Hospital, Capital Medical University, Beijing 100050, China

Correspondence: Yanwei Pang, School of Electrical and Information Engineering, Tianjin University, Tianjin 300072, China. Email: pyw@tju.edu.cn

**Purpose:** To introduce a dual-domain reconstruction network with V-Net and K-Net for accurate MR image reconstruction from undersampled k-space data.

**Methods:** Most state-of-the-art reconstruction methods apply U-Net or cascaded U-Nets in image domain and/or k-space domain. Nevertheless, these methods have following problems: (1) Directly applying U-Net in k-space domain is not optimal for extracting features in k-space domain; (2) Classical image-domain oriented U-Net is heavy-weight and hence is inefficient to be cascaded many times for yielding good reconstruction accuracy; (3) Classical image-domain oriented U-Net does not fully make use information of encoder network for extracting features in decoder network; and (4) Existing methods are ineffective in simultaneously extracting and fusing features in image domain and its dual k-space domain. To tackle these problems, we propose in this paper (1) an image-domain encoder-decoder sub-network called V-Net which is more light-weight for cascading and effective in fully utilizing features in the encoder for decoding, (2) a k-space domain sub-network called K-Net which is more suitable for extracting hierarchical features in k-space domain, and (3) a dual-domain reconstruction network where V-Nets and K-Nets are parallelly and effectively combined and cascaded.

**Results:** Extensive experimental results on the challenging fastMRI dataset demonstrate that the proposed KV-Net can reconstruct high-quality images and outperform current state-of-the-art approaches with fewer parameters.

**Conclusions:** To reconstruct images effectively and efficiently from incomplete k-space data, we have presented a parallel dual-domain KV-Net to combine K-Nets and V-Nets. The KV-Net is more lightweight than state-of-the-art methods but achieves better reconstruction performance.

**Word Count:** 4728

## 1 INTRODUCTION

Magnetic Resonance Imaging (MRI) is one of the most powerful diagnostic tools for a wide range of disorders. As a non-invasive, radiation-free, and in-vivo imaging modality, MRI can provide better soft tissue contrast than many other imaging modalities.

However, in the case of 2D Cartesian acquisition, MRI requires a long scan time to acquire a complete k-space complex matrix $\mathbf{K} = (\mathbf{k}_1,...,\mathbf{k}_P) \in \mathbb{C}^{F \times P}$ by encoding $F$ frequencies and $P$ phases so that the Nyquist sampling theorem is satisfied and the image can be perfectly reconstructed from $\mathbf{K}$ by Inverse Fast Fourier Transform (IFFT). Each column vector $\mathbf{k}_j \in \mathbb{C}^F$ can be called a phase vector. The scan time is proportional to the number (i.e., $P$) of acquired phase vectors). Usually, the scan time can range from 15 minutes to 90 minutes. The long scan time makes patients uncomfortable, leads to motion artifacts, and hampers fast diagnosis.

Therefore, it is of great importance to develop fast MRI where scan time is reduced [8,9,10,11,12] by partially encoding a small number (denoted by $C$ and $C<P$) of phase vectors. The acquired k-space matrix of such partial scan is expressed as $\bar{\mathbf{K}} \in \mathbb{C}^{F \times P}$ with $C$ columns being acquired phase vectors and the rest $P-C$ columns being zero-valued vectors. The incomplete matrix $\bar{\mathbf{K}}$ can also be called undersampled k-space matrix. The scan acceleration factor $AF$ is defined as $AF = P/C$. This paper focuses on developing algorithms for accurately reconstructing images from the undersampled k-space matrices acquired by a single coil. Extension to the case of parallel imaging where multi-coil data [7,13,15,16,21] is possible but is out of the scope of the paper.

Traditional compressed sensing methods [4,17,24,26,32,47,51] had made the breakthrough in reconstructing from undersampled MR images and/or incomplete k-space matrices. However, the past few years have witnessed remarkable superiority of deep learning based methods [3,5,19,22,25,27,33,34,37,40,42,45,58,59] over non-deep learning methods (e.g., classical Compressed Sensing methods [28,29]) when large-scale training data is available [6,18,20,51,53]. Therefore, this paper is concerned with deep learning based reconstruction methods.

As surveyed by Wang et al. [53], there are many kinds of deep learning based reconstruction methods. However, it is difficult to fairly compare the different methods on small isolated datasets compiled by independent groups and, in many cases, not shared with the greater research community [54]. The fastMRI dataset released in Dec. 2019 provides a good opportunity and a benchmark to compare these methods [51,54]. The training set and validation set of fastMRI dataset consist of large-scale raw k-space data and DICOM (Digital Imaging and Communications in Medicine) images. The released test data contains only the incomplete k-space matrices and undersampled images. It is noted that the ground truth (complete k-space matrices) of the test data is not released and the test results can only be observed in the public leaderboard [55] if the reconstructed images were properly uploaded.

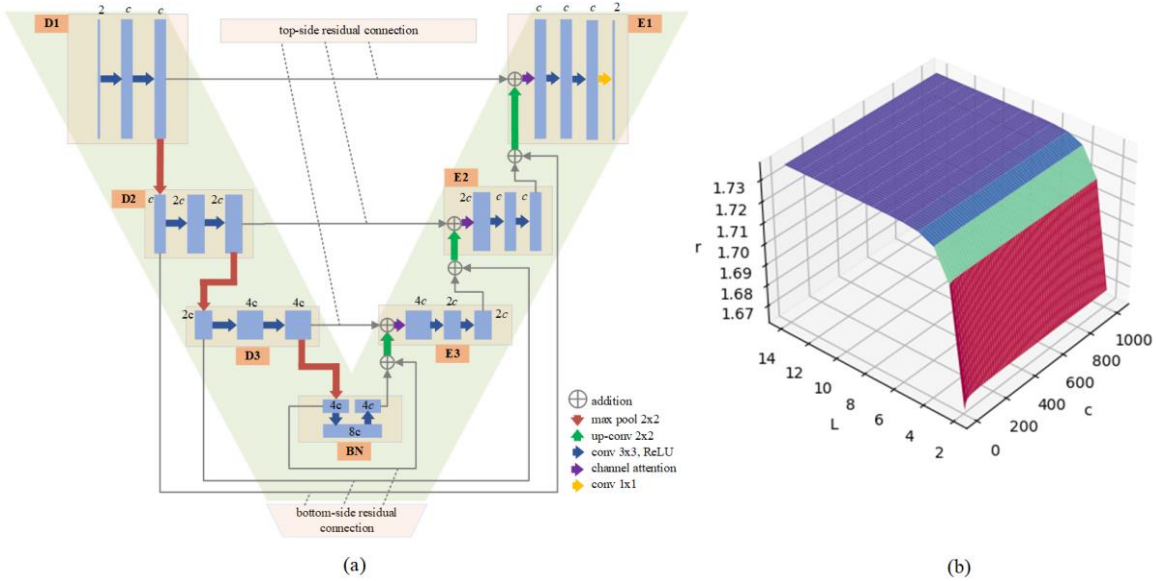

Figure 1. (a) The architecture of V-Net. A block D$i$ of the encoder is horizontally symmetric to the block E$i$ of the decoder. The top-side residual connection connects the tail of a block of the encoder and the head of the horizontally symmetric block in the decoder. The bottom-side residual connection connects the head of a block of encoder and the tail of the horizontally symmetric block in the decoder. The channel number of the last two layers of a block in the decoder is half of that of the first layer so that element-wise addition for residual connection is applicable. (b) The ratio r =$C_u/C_v$ between $C_u$ and $C_v$ varies with c and L.

It can be found from the leaderboard that most state-of-the-art methods (top-10 methods) employ U-Net [44] or its variants [2,23,33,57]. Here U-Net is the classical encoder-decoder form of Convolutional Neural Network (CNN). Despite great success, these methods are not optimal and have four problems: (1) Classical U-Net is originally designed for data in image space. Directly applying the image-oriented U-Net in k-space data is not optimal for extracting features in the frequency domain [31,60]; (2) Classical U-Net is heavy-weight and hence is inefficient to be cascaded many times for yielding good reconstruction accuracy [56]; (3) Classical U-Net does not fully make use information of encoder network for extracting features in decoder network [43,51]; and (4) Existing methods are ineffective in simultaneously extracting and fusing features in the image domain and its dual k-space domain [1,20].

This paper is aimed to tackle the above-mentioned problems by designing an effective dual-domain reconstruction network with its subnetworks being more suitable for MR image reconstruction. Here, dual-domain means image domain and its dual k-space (frequency) domain. The novelties and contributions of the paper are as follows.

(1) V-Net is proposed to overcome the limitations of classical U-Net. To decode high-resolution features, two-side residual connection between encoder and decoder is applied in V-Net whereas merely one-side concatenation connection is employed in U-Net [44] and traditional V-Net [30]. The proposed two-side connection is superior in not only utilizing features in the encoder but also reducing model size. The light-weightness of V-Net makes it more suitable for cascading networks with relatively small number of parameters.

(2) K-net is proposed to overcome the limitation of directly applying U-Net on k-space data. The novelties of the K-Net lie in the proposed cross-domain pooling (downsampling) and cross-domain upsampling. Compared with U-Net, the proposed K-Net is more specific to extracting hierarchical features in the frequency domain.

(3) A dual-domain reconstruction network, KV-Net, is proposed to parallelly and effectively integrate image domain oriented V-Nets and frequency domain oriented K-Nets. The parallel manner of dual-domain integration is in contrast to the existing sequential and alternating manner of dual-domain integration.

(4) The proposed method achieves competitive reconstruction accuracy with small number of parameters. Among the methods whose details were publicly available, the proposed method performs the best on the single-coil knee dataset of the challenging fastMRI dataset [55].

## 2 METHODS

We propose a parallel dual-domain framework, KV-Net, for reconstructing high-quality MR images with smaller number of parameters. We begin with the proposed V-Net and K-Net for overcoming the limitations of U-Net.

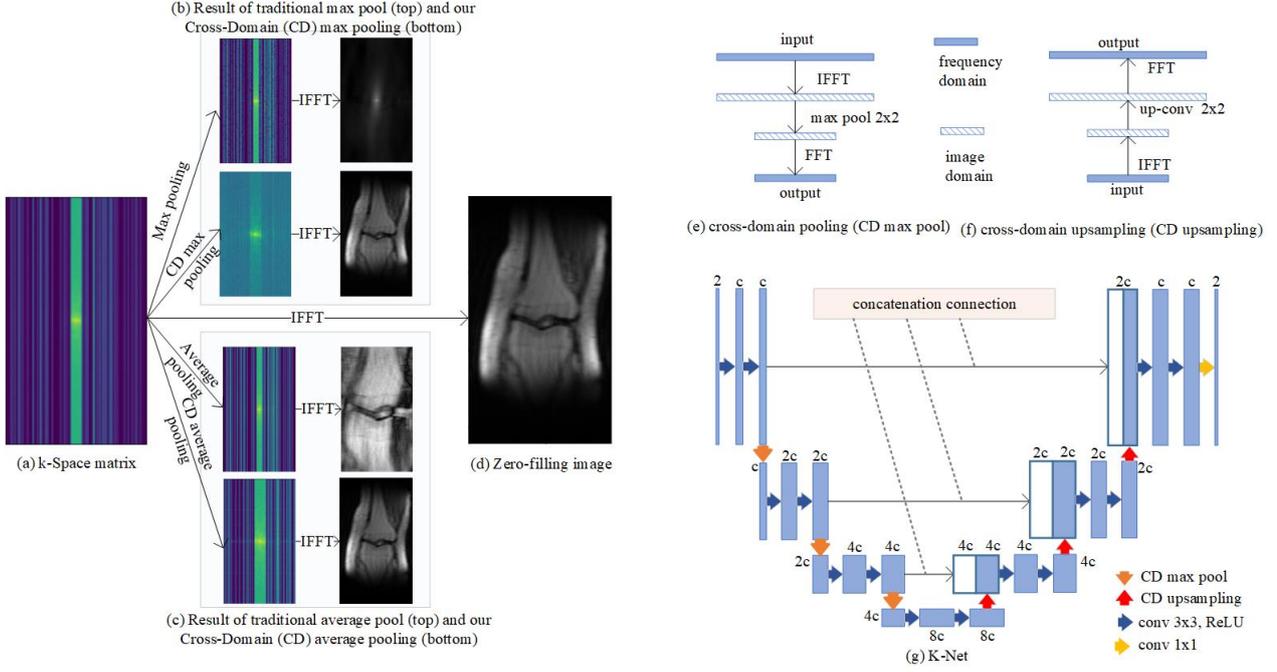

Figure 2. Traditional pooling is not suitable for frequency data (e.g., k-space matrix) and the architecture of K-Net. (a) Input a k-space matrix. (b) Result of traditional max pool (top) and our Cross-Domain (CD) max pooling (bottom). (c) Result of traditional average pool (top) and our Cross-Domain (CD) average pooling (bottom). (d) Zero-filling image. (e) Cross-domain pooling (CD max pooling). (f) Cross-domain upsampling (CD upsampling). (g) K-Net with CD max pooling and CD upsampling. The entry channel number c=8 is used.

## 2.1 PROPOSED LIGHT-WEIGHT IMAGE-DOMAIN V-NET

The proposed V-Net is targeted at solving the drawbacks of the classical U-Net [44]. Compared with U-Net, though the proposed V-Net is simple, it can yield better reconstruction accuracy when much smaller number of parameters are used.

As a classical encoder-decoder form of CNN, U-Net plays an important role in MR image reconstruction. In U-Net, the decoder network is strictly symmetric to the encoder network in the sense of structure, skip concatenation connection between encoder and decoder, and channel number. U-Net has the following two drawbacks which play negative roles for image reconstruction. (1) Existing skip connection is not able to fully adopt features in the encoder for the decoder to restore important details. Merely the last layer of a block of the encoder is connected to the first layer of the corresponding block of the decoder. (2) The symmetricity of channel number between encoder and decoder makes U-Net is not light-weight. When such U-Net is cascaded many times, the total number of parameters will be very large.

Figure 1a illustrates the proposed V-Net. In Figure 1a, there are three blocks (D1, D2, D3) in the encoder and three blocks (E1, E2, E3) in the decoder and there is a linear BottleNeck (BN) between D3 and E3. Block D$i$ and block E$i$ are horizontally symmetric for $i$=1, 2, …, $L$. $L$ stands for the number of blocks. It is noted that V-Net can be easily extended by changing the block numbers.

**Input and output**: The input is a two-channel undersampled image with one channel for the real part and the other for the imagery part of a complex number. Accordingly, the output also has two channels which are an estimation of fully sampled complex image (ground truth). In the first block, the input is transformed to $c$ channels by convolution followed by nonlinear activation. Throughout the paper, we call $c$ the entry channel number.

**Encoder:** Without considering how encoder and decoder are connected, the encoder of V-Net is the same as that of U-Net. The channel number of a block is double the channel number of the preceding block. Let $L$ be the number of blocks. Then parameter number $C_d$ of the encoder is:

$$C_d = (2 \times c + c \times c + \sum_{i=1}^{L-1} 2^{i-1} c \times 2^i c + 2^i c \times 2^i c) \times (3 \times 3) \tag{1}$$

**Two-side residual connection:** The differences between V-Net and U-Net lie in the decoder, BN, how the decoder and encoder are connected, and the results of the connection. As shown in Figure 1a, a top-side residual connection is used to connect the last layer of a block of the encoder and the first layer of the corresponding block of the decoder (the two blocks are horizontally symmetric). Moreover, a bottom-side residual connection is used to connect the first layer of a block of the encoder and the last layer of the corresponding block of the decoder (the two blocks are also horizontal symmetric). The introduced bottom-side symmetric connection benefits V-Net for utilizing more information of encoder for recovering details in the decoder [23]. By contrast, there is only one side connection in U-Net. There is also a bottom-side residual connection in BN of V-Net which is different from U-Net.

**Parameter reduction in decoder and bottleneck:** Because element-wise addition of residual connection can be conducted only when the channel numbers of the two layers to be connected are the same, the channel number of the last two layers is required to be half of the channel number of the first layer in the same decoder block. For example, the channel number of block E3 is 4c and the channel number of the last two layers of E3 is 2c. The parameter number, $C_e$, of the convolutional layers in the decoder is:

$$C_e = 2 \times c + (2 \times c \times c + \sum_{i=1}^{L-1} 2^i c \times 2^{i-1} c + 2^{i-1} c \times 2^{i-1} c) \times (3 \times 3) \qquad (2)$$

To adopt the bottom-side residual connection in BN, the channel number of the last layer is reduced to be half of the channel number of the second layer in BN. Then parameter number $C_{bn}$ of the BN is:

$$C_{bn} = (2^{L-1} c \times 2^L c + 2^L c \times 2^{L-1} c) \times (3 \times 3) \qquad (3)$$

The parameter number of the transposed convolution layers are also reduced with the change of $C_e$ and $C_{bn}$:

$$C_{up} = (\sum_{i=1}^{L} 2^{i-1} c \times 2^{i-1} c) \times (2 \times 2) \qquad (4)$$

Totally, for an $L$ level V-Net, the parameter number is $C_v$:

$$C_v = C_d + C_e + C_{bn} + C_{up} \qquad (5)$$

For an $L$ level U-Net, the parameter number is $C_u$:

$$C_u = C_d + 2 \times c + (\sum_{i=1}^{L} 2^i c \times 2^{i-1} c + 2^{i-1} c \times 2^{i-1} c) \times (3 \times 3)$$
$$+ (2^{L-1} c \times 2^L c + 2^L c \times 2^L c) \times (3 \times 3) \qquad (6)$$
$$+ (\sum_{i=1}^{L} 2^i c \times 2^{i-1} c) \times (2 \times 2).$$

The ratio $r = C_u/C_v$ between $C_u$ and $C_v$ is a function of $c$ and $L$. The curves of $r$ versus $c$ and $L$ are shown in Figure 1b. For the case of $c=32$ and $L=3$, $r=1.72$ holds. That is, the model size of V-Net is typically 1.72 times smaller than U-Net.

Therefore, compared with U-Net, V-Net is superior in not only utilizing features in the encoder but also reducing model size. The light-weightiness of V-Net makes it more suitable for cascading networks with relatively small number of parameters.

**Channel attention for top-side residual connection:** We propose to apply classical channel attention, Squeeze and Excitation (SE) attention [48] on the result of top-side residual connection so that each channel is adaptively weighted. In Figure 1a, channel attention is applied in Blocks E1, E2, and E3. The parameter number and computational cost of the channel attention are negligible compared with those of the whole model.

## 2.2 PROPOSED K-SPACE DOMAIN K-NET

Generally speaking, neighboring pixels of an image are closely correlated hence can be well modeled but neighboring elements of a single-coil k-space matrix are not so correlated and so is difficult to be modeled. Nevertheless, it is known that directly modeling the k-space data with a neural network is capable of providing information unable to be provided by image-domain modeling [6]. Therefore, it is important to perform MR image reconstruction in the k-space domain (frequency domain).

**Cross-domain downsampling:** Most existing methods directly apply image-oriented neural networks (e.g., U-Net) on the k-space domain [14, 31, 60]. However, we discover that U-Net, originally designed for the image domain, is not optimal for the k-space/frequency domain. Specifically, the operation of max-pooling (or other pooling) and upsampling in U-Net are not suitable for the frequency domain.

Examples of failures of traditional max-pooling and average pooling are shown in Figure 2. Figure 2a is the given input of a k-space matrix. The results of the traditional max-pooling are shown in Figure 2b. From the image version (obtained by applying IFFT on the pooled result) of the result, it can be seen that applying traditional max-pooling results in a significant drop in results, where the appearance of the slice cannot be observed at all. The results of the traditional average pooling are shown in Figure 2c, the image version is enlarged and a lot of image information is lost. Analogously, traditional upsampling used in the decoder is also not optimal for frequency-domain data.

To solve the problem of traditional pooling methods, we propose a frequency-domain oriented pooling method which we call Cross-Domain (CD) pooling. Figure 2b and 2c show exemplary results of the proposed CD pooling. It can be seen from Figure 2b that the CD max-pooling is much better than the traditional max-pooling, the CD average pooling is also much better than the traditional average pooling in Figure 2c.

The proposed CD pooling module is illustrated in Figure 2e. The input (a layer of feature maps) to be downsampled is firstly transformed to the image domain by IFFT. Then traditional pooling (e.g., max-pooling, average pooling, and stride-2 convolution) is applied to obtain downsampled results in the image domain. Finally, the image version of the downsampled layer is transformed to the frequency domain by FFT.

**Cross-domain upsampling:** Analogously, we propose a frequency-domain oriented upsampling method which we call Cross-Domain (CD) upsampling (Figure 2f). As shown in Figure 2f, CD upsampling can be viewed as an inversion of CD downsampling. The input (a layer of feature maps) to be upsampled is firstly transformed to the image domain by IFFT. Then traditional upsampling (e.g., bilinear interpolation and deconvolution) is applied to obtain upsampled results in the image domain.

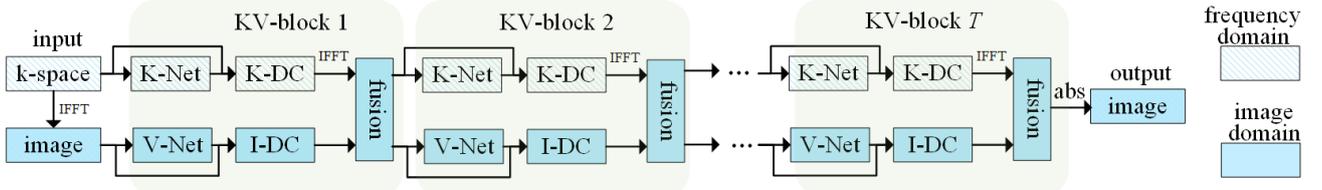

Figure 3. Architecture of KV-Net. $T$ KV-blocks are cascaded. A K-Net and a V-Net in a KV-block are parallelly fused. In each KV-block, the outputs of the K-Net and the V-Net are processed by k-space Data-Consistency (K-DC) and image-space Data-Consistency (I-DC). abs computes the magnitude of the output (complex-valued matrix) of KV-block $T$. The entry channel number of K-Net and V-Net are denoted by $c_k$ and $c_v$, respectively.

Finally, the image version of the upsampled layer is transformed to the frequency domain by FFT.

**Lightweight K-Net:** Based on the cross-domain pooling and the cross-domain upsampling, we construct a k-space oriented network which we call K-Net. As shown in Figure 2g the architecture of K-Net is almost identical to that of traditional U-Net. The differences lie in the pooling and upsampling operators. Traditional pooling and upsampling are replaced with cross-domain pooling and cross-domain upsampling, respectively. In our implementation, the entry channel number $c$ is set to 8 so that the K-Net is lightweight. The lightweight K-Net with small $c$ almost performs as well as a heavy-weight U-Net with large $c$.

## 2.3 PARALLEL DUAL-DOMAIN KV-NET

Because K-Net and V-Net can extract complementary features, we propose a KV-Net to combine K-Nets and V-Nets to reconstruct high-quality images.

### 2.3.1 Overall Architecture

The architecture of KV-Net is shown in Figure 3. The input is incomplete k-space matrices and the corresponding image-space undersampled images. In contrast to the manner of alternating between a k-space network and an image-domain network, we propose to construct a parallel dual-domain block, called KV-block, by two parallel branches: a frequency-domain K-Net branch and an image-domain V-Net branch. Predefined numbers (denoted by $T$) of KV-blocks are sequentially cascaded. The entry channel number of K-Net and V-Net are denoted by $c_k$ and $c_v$, respectively.

In each KV-block, the outputs of the K-Net and the V-Net are processed by k-space Data-Consistency (K-DC) and image-space Data-Consistency (I-DC). The output of K-DC is transformed to the image domain by IFFT and then is fused with the output of I-DC. The fusion result and its frequency-domain version (obtained by applying FFT on the fusion result) are viewed as input to the next KV-block. The output of the last KV-block (i.e., KV-block $T$), consisting of real parts and imagery parts is finally converted to a reconstructed image by computing the magnitudes of real and imaginary parts. The output of the KV-Net is supervised by the ground truth of fully sampled images obtained by applying IFFT on the complete k-space matrices.

The data-consistency module, fusion module, and the loss function are described as follows.

### 2.3.2 Data Consistency Modules: K-DC and I-DC

The data consistency module plays a vital role in the whole cascaded framework. It is used to impose the constraints from the raw MR measurement and improve data fidelity. The softer DC [49] is adopted in K-DC:

$$k_{rec}(j) = \begin{cases} \hat{k}(j) & if\ j \notin \Omega, \\ \hat{k}(j) - \gamma(\hat{k}(j) - k(j)) & if\ j \in \Omega. \end{cases} \quad (7)$$

In Eq. (7), $j$ represents the index of the vectorized representation of k-space data, $\hat{k}$ denotes the output of a K-Net, $k$ denotes the raw incomplete k-space data, $\Omega$ is the index set of sampling data, and $\gamma$ is a trainable hyperparameter to balance predicted data and raw sampling data.

As defined in (7), if the position of the input k-space data point is not sampled ($j \notin \Omega$), the predicted values are used directly. For the sampled points, a trainable parameter $g$ is used to punish the distance from the raw measured value to balance the consistency and smoothness of the data.

I-DC is an image-domain version of K-DC. The output of V-Net is transformed to its frequency domain by FFT. Then the softer DC expressed in Eq. (7) is applied. Finally, the result of DC is transformed back to the image domain by IFFT.

### 2.3.3 Fusion Module

The output (denoted by $B_k$) of K-Net followed by K-DC is in the frequency domain and the output (denoted by $A_v$) of V-Net followed by I-DC is in the image domain. The fusion module is to combine the two different outputs. We propose to fuse them in image domain. Firstly, $B_k$ is transformed to $A_k$ by $A_k$=IFFT($B_k$). Then $A_k$ and $A_v$ are linearly combined:

$$A = \frac{1}{1+\mu}A_v + \frac{\mu}{1+\mu}A_k \quad (8)$$

where $A$ is the result of fusion and $\mu$ is a learnable balancing coefficient.

### 2.3.4 Loss Function

The output (denoted by $\hat{x}$) of the KV-Net is supervised by ground truth (denoted by $x$) of fully sampled images obtained by applying IFFT on the complete k-space matrices. The following Structural Similarity (SSIM)[50] loss function $J(\hat{x}_0, x)$ is used to train the KV-Net:

$$J(\hat{x}_0, x) = 1 - SSIM(\hat{x}_0, x) \qquad (9)$$

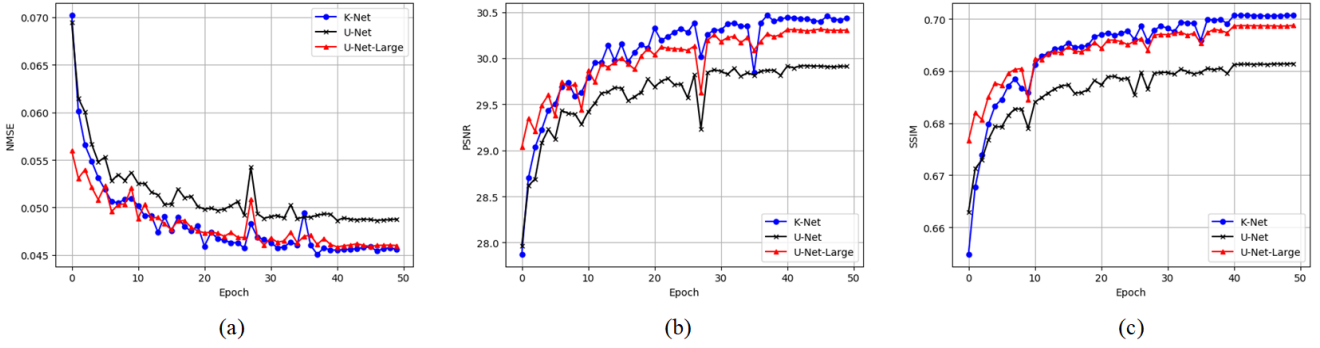

Figure 4. Comparison of K-Net with k-space domain U-Net on validation set. The input of K-Net and U-Net is a k-space incomplete matrix. The architectures of the K-Net and U-Net are identical except the pooling and upsampling operation. Cross-domain pooling and cross-domain upsampling are employed in K-Net. $c$ is the entry channel number.

In Eq. (9), $SSIM(\hat{x}_0, x)$ is the value of SSIM between $\hat{x}_0$ and $x$.

## 3 RESULTS

### 3.1 EXPERIMENTAL SETTINGS

**Dataset** The proposed method is evaluated on the fastMRI single-coil knee dataset[51]. The dataset, acquired by 3T and 1.5T MRI devices, contains both raw k-space data and DICOM images. The dataset is organized as a lot of volumes with each volume roughly consisting of 36 slices. The training set, validation set, and test set consist of 973, 199, and 108 volumes. The ground truth of the test data has not been released.

**Cartesian undersampling method** In all the experiments, the undersampled images of training and validation sets are obtained by the $4\times$ acceleration random undersampling masks. The source codes released with the fastMRI dataset are employed to generate undersampling masks. The undersampling masks include a fully-sampled central region, taking 8% of all k-space lines for the $4\times$ acceleration. The remaining k-space lines are included uniformly at random to achieve desired acceleration factor on average.

**Training setting** The RMSProp optimizer is used to optimize the network. The initial learning rate is 0.001 and will decrease to 0.0001 at 40 epochs. The PyTorch framework is adopted for model implementation. A machine with 4 NVIDIA Tesla V100 GPUs is used for data parallel training.

**Performance Evaluation** Averaged Structural Similarity (SSIM), Peak Signal to Noise Ratio (PSNR), and Normalized Mean Squared Error (NMSE) are adopted to measure the quality of images reconstructed[50, 51]. The performance for reconstructing test data is available on the fastMRI Single-Coil Knee Public Leaderboard[55].

The proposed KV-Net is compared with the following six state-of-the-art algorithms on the fastMRI leaderboard: TV model[51], KIKI-Net[6], MD-Recon-Net[20], U-Net Baseline Model[51], XPDNet[52], and i-RIM[36]. The source codes of the KIKI-Net, MD-Recon-Net and U-Net baseline are downloaded from the websites of the authors of these methods or implemented strictly according to the original papers. The results of the other three methods are obtained directly from the fastMRI public leaderboard uploaded by the original authors.

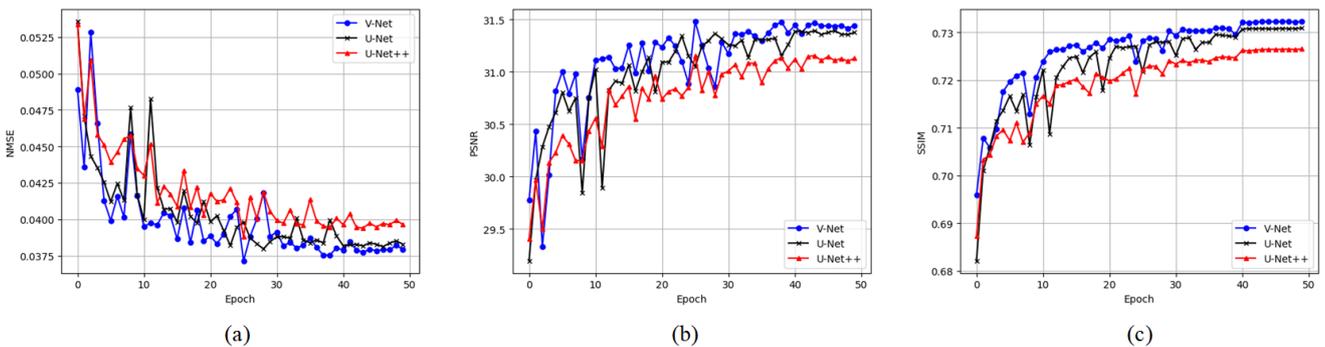

Figure 5. Comparison of V-Net with image-domain U-Net and U-Net++ on validation set. The input to V-Net, U-Net and U-Net++ is an undersampled image.

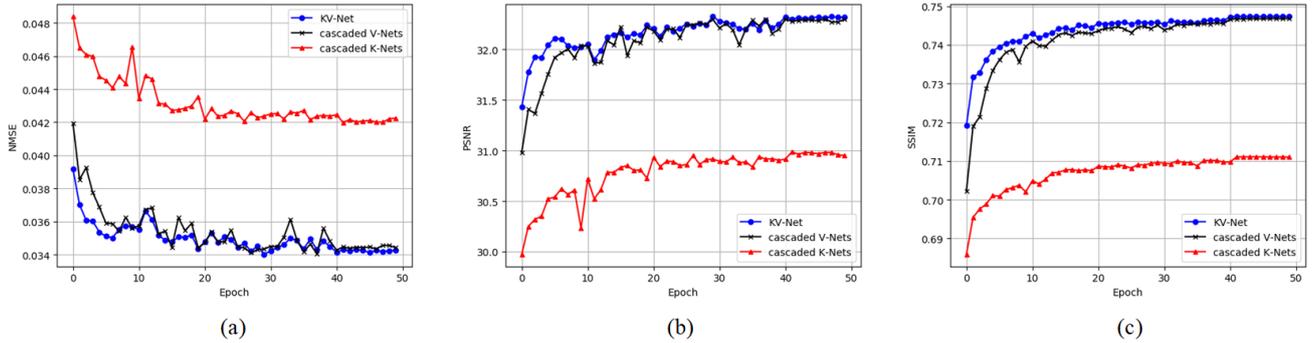

Figure 6. Comparison of KV-Net, cascaded K-Nets, and cascaded V-Nets on validation set. (a)~(c) shows how NSME, PSNR, and SSIM vary with epoch number, respectively.

## 3.2 ABLATION STUDIES

The proposed k-space K-Net, image-space V-Net, and dual-domain KV-Net are evaluated on the validation set.

### 3.2.1 Effectiveness of K-Net, cross-domain pooling and upsampling

The proposed K-Net is designed for directly reconstructing from the incomplete k-space matrices. Traditional U-Net can also be used for reconstructing from the incomplete k-space matrix. The difference between K-Net and the U-Net applied in the k-space domain lies in the proposed cross-domain pooling and cross-domain upsampling. Therefore, we compare the K-Net with the U-Net, which is the same as K-Net except for the pooling and upsampling operation. Figure 4 shows curves of SSIM, PSNR, and NSEM when the entry channel number is $c$=8. The horizontal axis stands for training epoch number. Table 1 gives the results when K-Net and U-Net converge after 50 epochs. The NSME, PSNR, and SSIM of K-Net are 0.0457, 30.41, and 0.7007, respectively. The NSME, PSNR, and SSIM of U-Net with $c$=8 are 0.0487, 29.91, and 0.6914, respectively. It is observed from Figure 4 that K-Net consistently outperforms U-Net, implying great importance of the proposed cross-domain pooling and upsampling.

Table 1 also gives the results of U-Net-Large with the entry channel number being $c$=32 and parameter number being 1.9 M (Million). Even though the proposed lightweight K-Net has as small as 0.1M parameters, it outperforms heavy-weight U-Net-Large in terms of NSME, PSNR, and SSIM.

### 3.2.2 Effectiveness of V-Net

The proposed V-Net is designed for reconstructing from undersampled images. Figure 5 and Table 1 compare the V-Net with the image-domain U-Net and U-Net++ whose input are also undersampled images and entry channel numbers are equal. Figure 5 shows how reconstruction performances (i.e., SSIM, NSME, and PSNR) vary with the epoch number. Table 1 gives parameter numbers and reconstruction performance when V-Net, U-net and U-Net++ converge. It can be seen that compared with U-Net and U-Net++, the proposed lightweight V-Net achieves better reconstruction quality even though V-Net has much smaller number of parameters. Specifically, when the entry channel number c is 32, V-Net, U-Net and U-Net++ have 1.1 M, 1.9 M and 2.2 M parameters, respectively. In this situation, the NSME, PSNR, SSIM of V-Net are respectively 0.0379, 31.44, and 0.7323, whereas the NSME, PSNR, and SSIM of U-Net are 0.0382, 31.39, and 0.7307, and the NSME, PSNR, and SSIM of U-Net++ are 0.0397, 31.12, and 0.7265.

### 3.2.3 Effectiveness of combing K-Nets and V-Nets

As shown in Figure 3, KV-Net has a cascaded K-Net branch and a cascaded V-Net branch. To independently investigate the role of each branch and effectiveness of combining the two branches, the K-Net branch is removed from the KV-Net, and the resultant network is called cascaded V-Nets. Similarly, the V-Net branch is removed from the KV-Net, and the resultant network is called cascaded K-Nets. The entry channel number of K-Net and V-Net is set to $c_k$=8 and $c_v$=32, respectively. The number of KV blocks is set to $T$=12.

The results are shown in Figure 6 and Table 1. Figure 6 shows how NSME, PSNR, and SSIM vary with the epoch number. Table 1 gives the results when the algorithms converge after training 50 epochs. The SSIM of cascaded V-Nets, cascaded K-Nets, and KV-Net is 0.7468, 0.7111, and 0.7474, respectively. The PSNR of cascaded V-Nets, cascaded K-Nets, and KV-Net is 32.30, 30.98, and 32.32, respectively. The NSME of cascaded V-Nets, cascaded K-Nets, and KV-Net is 0.0344, 0.0420, and 0.0342, respectively.

It is concluded from the results that combing cascaded V-Nets and cascaded K-Nets to KV-Net improves the reconstruction quality significantly, and the proposed parallel dual-domain fusion is effective.

## 3.3 COMPARISON WITH STATE-OF-THE-ART METHODS

The proposed KV-Net is compared with TV model [51], KIKI-Net [6], MD-Recon-Net [20], U-Net Baseline Model [51], XPDNet [52], and i-RIM [36] on the test data of fastMRI dataset. In addition, we compare KV-Net with MD-Recon-Net, U-Net baseline, and KIKI-Net on validation set because the source codes of these methods are available and can produce the curves of reconstruction quality versus epoch number.

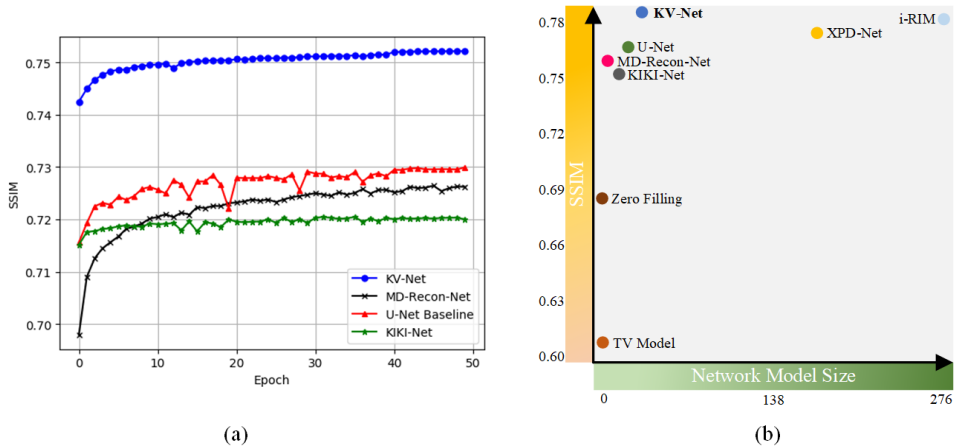

Figure 7. Validation and test set results. (a) Comparison of KV-Net, MD-Recon-Net [20], U-Net baseline [51], and KIKI-Net [6] on validation set. The curves of SSIM versus epoch number are shown. In KV-Net, the number of KV blocks is set to T=12. (b) The curves of SSIM versus model size (M). The proposed KV-Net gets the best balance between SSIM and model size.

Hereinafter, the entry channel number of K-Net and V-Net are $c_k$=8 and $c_v$=32, respectively, $L$=3. The number of KV blocks is $T$=12.

### 3.3.1 Comparison on validation set

Figure 7a shows the reconstruction results of KV-Net, MD-Recon-Net, U-Net baseline, and KIKI-Net on the validation set. From the first epoch to the last epoch, the SSIM of KV-Net is always remarkably higher than that of other methods. Specifically, the final SSIM of KV-Net, MD-Recon-Net, U-Net baseline, and KIKI-Net are 0.7521, 0.7262, 0.7299, and 0.7199, respectively. Here, final SSIM means the SSIM when a reconstruction algorithm converges.

### 3.3.2 Comparison on test set

Almost all the ground truth of the test set is not accessible because it is not released. Only several ground truth images are shown for visual comparison once the reconstructed images of the whole test data are uploaded to the fastMRI official website. The SSIM, NSME, PSNR of the reconstructed images are reported on the public leaderboard [55].

Table 2 gives the results of the proposed KV-net and other state-of-the-art methods on the test data. SSIM is the most important criterion for Image Quality Assessment (IQA) because of its consistency with human visual systems. Therefore, we mainly analyze the SSIM results of different methods. The SSIM of Zero Filling, TV model [51], KIKI-Net [6], MD-Recon-Net [20], U-Net Baseline Model [51], XPDNet [52], i-RIM [36], and the proposed KV-Net is 0.6870, 0.6028, 0.7520, 0.7590, 0.7604, 0.7763, 0.7807, and 0.7814, respectively. The proposed KV-Net achieves the largest SSIM. Among the eight methods, only i-RIM and KV-Net are capable of producing SSIM larger than 0.7800. But i-RIM has 275$M$ parameters, whereas there are as small as 14$M$ parameters in the proposed method. Compared with i-RIM, KV-Net can reconstruct more accurate images with one-order fewer parameters.

The superiority of KV-Net over Zero Filling, TV model, KIKI-Net, MD-Recon-Net, U-Net Baseline Model, XPDNet, and i-RIM can also be intuitively seen from Figure 7b where the curves of SSIM versus model size (i.e., number of parameters of a model) are shown. It is ideal if a method gets the largest SSIM with the smallest number of parameters. That is, the tuple (SSIM, model size) lies in the most up-left corner of Figure 7b. Though there is no such an ideal method, the proposed method, KV-Net, achieves the best balance between reconstruction quality (in the sense of SSIM) and model size.

The NSME and PSNR of KV-Net are equal to those of the heavy-weight i-RIM and are better than that of the U-Net baseline and MD-Recon-Net. XPD-Net gets impressive results in terms of PSNR and NSME. But its SSIM is below 0.78 and has 155 $M$ parameters. By contrast, the SSIM and parameter number of KV-Net are 0.7814 and 14 $M$.

As shown in Figure 8, the merit of KV-Net can also be observed from reconstructed images. Compared with other methods, the appearance and details of the image reconstructed by the proposed KV-Net are the best to approximate the appearance and details of the ground truth.

Specifically, the analysis of Figure 8 is as follows. There are severe artifacts in images reconstructed by Zero Filling and TV Model. The non-deep-learning methods, Zero Filling and TV Model, are obviously inferior to the deep-learning based methods. As to deep-leaning based methods, U-Net Baseline outperforms KIKI-Net and MD-Recon-Net in suppressing artifacts. But there are stubborn artifacts in the results of U-Net Baseline. In addition, U-Net Baseline tends to produce over-smoothed images because only single-domain (i.e., image-domain) information is utilized in U-Net Baseline. The images reconstructed by XPD-Net, i-RIM, and KV-Net are significantly better than that of U-Net. KV-Net outperforms XPD-Net and i-RIM in suppressing artifacts and preserving textures and details in high-frequency bands.

The merits of KV-net are owed to the effective and low-weight image-domain V-Net, cross-domain pooling and upsampling in K-Net, and parallel dual-domain structure for fusion of V-Nets and K-Nets.

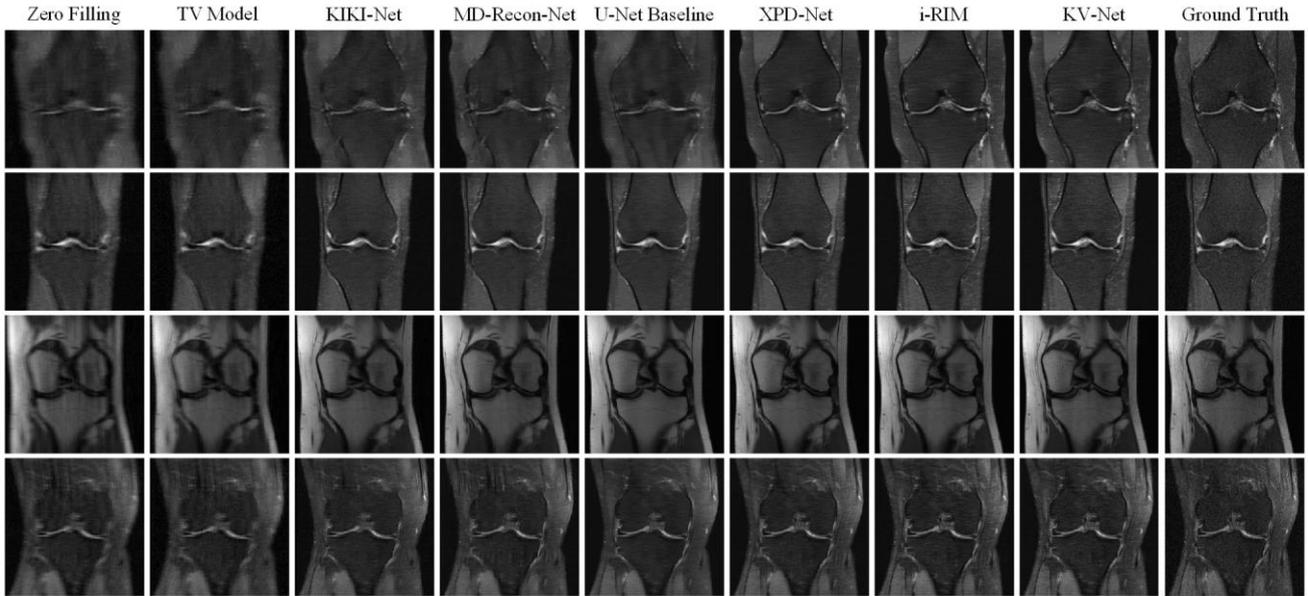

Figure 8. Examples of reconstructed images and the ground truths, available at the public leaderboard of fastMRI [55].

## 4 DISCUSSION AND CONCLUSION

To reconstruct images effectively and efficiently from incomplete k-space data, we have presented a k-space oriented K-Net, image-domain oriented V-Net, and parallel dual-domain KV-Net to combine K-Nets and V-Nets. K-Net is able to overcome the drawback of directly applying classical U-Net on the k-space domain by the proposed cross-domain pooling and cross-domain upsampling. In the image domain, V-Net is able to improve U-Net by the proposed two-side residual connection and parameter reduction in the sense of fully utilizing information of encoder for extracting expressive features in the decoder. In KV-Net, several KV-blocks are cascaded. In each KV-block, there is an image-domain branch (U-Net branch) parallel to the k-space branch (K-Net branch). The parallel manner of fusing K-Nets and V-Nets in the image domain is effective. Because V-Net is more lightweight than U-Net and K-Net is much more low-light than V-Net, the KV-Net is lightweight compared with state-of-the-art methods. Extensive experimental results on the challenging fastMRI dataset demonstrated that the proposed method could reconstruct high-quality images with a small number of parameters. In the future, we plan to extend KV-Net to reconstruct MR images from multi-coil data.

## ACKNOWLEDGEMENT

We would like to thank Dr. Jinghua Wang for correcting typos of the manuscript and Mr. Yong Sun for running the experiment about U-Net++.

TABLE 1 Ablation studies results. Ablation study 1 is the comparison results of K-Net with k-space domain U-Net on validation set. Ablation study 2 is the comparison results of V-Net with image-domain U-Net and U-Net++ on validation set. Ablation study 3 is the comparison results of of KV-Net, cascaded K-Nets, and cascaded V-Nets on validation set. $L$=3 in all the ablation studies.

| Ablation Studies | Method | Input | $c$ | Parameters | NMSE↓ | PSNR↑ | SSIM↑ |
|---|---|---|---|---|---|---|---|
| 1 | U-Net-Large | k-space | 32 | 1.9 M | 0.0460 | 30.30 | 0.6986 |
| 1 | U-Net | k-space | 8 | 0.1 M | 0.0487 | 29.91 | 0.6914 |
| 1 | **K-Net** | k-space | 8 | **0.1 M** | **0.0457** | **30.41** | **0.7007** |
| 2 | U-Net | image | 32 | 1.9 M | 0.0382 | 31.39 | 0.7307 |
| 2 | U-Net++ [57] | image | 32 | 2.2 M | 0.0397 | 31.12 | 0.7265 |
| 2 | **V-Net** | image | 32 | **1.1 M** | **0.0379** | **31.44** | **0.7323** |
| 3 | cascaded V-Nets | image | 32 | 13.2 M | 0.0344 | 32.30 | 0.7468 |
| 3 | cascaded K-Nets | k-space | 8 | 1.2 M | 0.0420 | 30.98 | 0.7111 |
| 3 | **KV-Net** | image&k-space | 32&8 | 14.4 M | **0.0342** | **32.32** | **0.7474** |

TABLE 2 Comparison with the best models on the fastMRI leaderboard [55].

| Method | Parameters | NMSE↓ | PSNR↑ | SSIM↑ |
|---|---|---|---|---|
| Zero Filling | - | 0.0438 | 30.5 | 0.6870 |
| TV model [51] | - | 0.0479 | 30.7 | 0.6028 |
| KIKI-Net [6] | 1.25 M | 0.0296 | 32.8 | 0.7520 |
| MD-Recon-Net [20] | 0.3 M | 0.0272 | 33.3 | 0.7590 |
| U-Net Baseline [51] | 7.8 M | 0.0271 | 33.2 | 0.7604 |
| XPD-Net [52] | 155 M | 0.0251 | 33.9 | 0.7763 |
| i-RIM [36] | 275 M | 0.0271 | 33.7 | 0.7807 |
| **Our KV-Net** | 14 M | 0.0271 | 33.7 | **0.7814** |

Figure 1. (a) The architecture of V-Net. A block D$i$ of the encoder is horizontally symmetric to the block E$i$ of the decoder. The top-side residual connection connects the tail of a block of the encoder and the head of the horizontally symmetric block in the decoder. The bottom-side residual connection connects the head of a block of encoder and the tail of the horizontally symmetric block in the decoder. The channel number of the last two layers of a block in the decoder is half of that of the first layer so that element-wise addition for residual connection is applicable. (b) The ratio r =Cu/Cv between Cu and Cv varies with c and L.

Figure 2. Traditional pooling is not suitable for frequency data (e.g., k-space matrix) and the architecture of K-Net. (a) Input a k-space matrix. (b) Result of traditional max pool (top) and our Cross-Domain (CD) max pooling (bottom). (c) Result of traditional average pool (top) and our Cross-Domain (CD) average pooling (bottom). (d) Zero-filling image. (e) Cross-domain pooling (CD max pooling). (f) Cross-domain upsampling (CD upsampling). (g) K-Net with CD max pooling and CD upsampling. The entry channel number c=8 is used.

Figure 3. Architecture of KV-Net. $T$ KV-blocks are cascaded. A K-Net and a V-Net in a KV-block are parallelly fused. In each KV-block, the outputs of the K-Net and the V-Net are processed by k-space Data-Consistency (K-DC) and image-space Data-Consistency (I-DC). abs computes the magnitude of the output (complex-valued matrix) of KV-block $T$. The entry channel number of K-Net and V-Net are denoted by $c_k$ and $c_v$, respectively.

Figure 4. Comparison of K-Net with k-space domain U-Net on validation set. The input of K-Net and U-Net is a k-space incomplete matrix. The architectures of the K-Net and U-Net are identical except the pooling and upsampling operation. Cross-domain pooling and cross-domain upsampling are employed in K-Net. $c$ is the entry channel number.

Figure 5. Comparison of V-Net with image-domain U-Net and U-Net++ on validation set. The input to V-Net, U-Net and U-Net++ is an undersampled image.

Figure 6. Comparison of KV-Net, cascaded K-Nets, and cascaded V-Nets on validation set. (a)~(c) shows how NSME, PSNR, and SSIM vary with epoch number, respectively.

Figure 7. Validation and test set results. (a) Comparison of KV-Net, MD-Recon-Net [20], U-Net baseline [51], and KIKI-Net [6] on validation set. The curves of SSIM versus epoch number are shown. In KV-Net, the number of KV blocks is set to T=12. (b) The curves of SSIM versus model size (M). The proposed KV-Net gets the best balance between SSIM and model size.

Figure 8. Examples of reconstructed images and the ground truths, available at the public leaderboard of fastMRI [55].